\documentclass[11pt]{article}

\usepackage[top=50mm, bottom=50mm, left=50mm, right=50mm]{geometry}

\usepackage{lineno}
\usepackage{amssymb}
\usepackage{amsmath}
\usepackage{amsthm}
\usepackage{epsfig}
\usepackage{graphicx}
\usepackage{graphics}
\usepackage{float}
\usepackage{subfigure}
\usepackage{multirow}
\usepackage{color}
\usepackage{lineno}
\usepackage{fullpage}
\usepackage[normalem]{ulem} 
\usepackage{makeidx}
\usepackage{xspace}
\usepackage{wrapfig}

\makeindex

\begin{document}

\title{Vortex dynamics and applications to gaseous optical elements}

\author{D. Kaganovich, B. Hafizi, L. A. Johnson, and D. F. Gordon \\
Plasma Physics Division, Naval Research Laboratory, Washington, DC 20375}

\maketitle
\setcounter{page}{1}

\begin{abstract}
Experimental studies of the optical properties of compressible, viscous and rapidly-rotating gas flows (vortices) are presented.  Gas vortices can function as optical elements such as lenses or waveguides. The optical properties are determined from direct interferometric phase measurements and beam propagation analysis. Output beams are analyzed in terms of Zernike polynomials for a range of gas flow parameters, including choked flow. The absolute radial gas density distribution is measured and a technique for adjusting it is demonstrated. 
\end{abstract}

\section{Introduction}
Gas flow in the form of a rotation about an axis is known as a vortex. A typical vortex consists of two distinct regions: 1) an inner, or core region located near and around the principal axis, where the gas is locked into rigid body rotation (forced vortex) and 2) an outer region, where the fluid elements, while moving along a circular path, do not rotate about their own axis (free or potential vortex) \cite{Rath}. The transition between these two regions depends on details of the fluid and the surroundings. Numerous models have been proposed to describe the flow characteristics as a function of the vortex parameters \cite{Rankine, Burgers, Sullivan, Batchelor, Vatistas, Rodriguez}. 

In previous work we experimentally demonstrated \cite{Kaganovich_18} and simulated \cite{Johnson} the lensing properties of spinning gases. We showed that near the axis of rotation the gas density profile (and refractive index) has a parabolic radial dependence. Farther away from the axis the gas density deviates from a parabola, leading to aberrations. In a simple axisymmetric case it is possible to show analytically that rigid body rotation of the core has a parabolic density profile (Appendix A). By limiting the optical aperture to the core the aberrations can be substantially reduced.

In this paper we report results of direct measurements of the gas density distribution inside a vortex. Two distinct cases are described in detail: 1) short vortex with diameter of the core comparable to the length of the vortex (hereinafter referred to as the gas lens configuration), and 2) a vortex that is much longer than its core diameter. The latter configuration can potentially be operated as laser guiding structure (waveguide) if the gas is ionized (hereinafter referred to as the guiding structure configuration).

A schematic of the gas lens is depicted in Figure \ref{fig1}a. Compressed gas (red arrows) flows into the lens and is forced into rotation inside the lens body. Spinning gas exits the gas lens along its optical axis (blue arrows). The diameter of the exit holes defines the clear aperture of the lens.  In most cases here it is larger than the core diameter of the vortex. Gas supply lines and gas lens geometry were designed to allow for the possibility of choked flow at the exit holes of the lens. In all the experiments the internal diameter of the lens body was 5 or 6 mm, the exit holes were in a range of 1.5-2 mm in diameter, and the lens thickness varied between 1 and 2 mm. The guiding structure configuration had similar dimensions except the length along the optical axis (thickness in case of a lens) that was 16 mm.

\section{Experiments}
\label{sec:experiemnt}

The lens had two independent gas inlets. Gas flow into each inlet was controlled and/or measured by a mass-flow-rate controller. Pressure regulators upstream of the mass flow controllers regulated the stagnation pressure $p_0$ of the system. A high capacity vacuum pump evacuated gas from the vacuum chamber, an adjustable valve controlled the pumping rate, and the backing pressure $p_b$ was monitored by a vacuum gauge. The gas lines and the optical setup are shown schematically in Figure \ref{fig2}a. All measurements presented in this paper were done under steady state conditions, when the gas flow rate and vacuum chamber backing pressure were constant. Nitrogen gas was used in all experiments.

\begin{figure}[htbp]
\centering
\fbox{\includegraphics[width=0.5\linewidth]{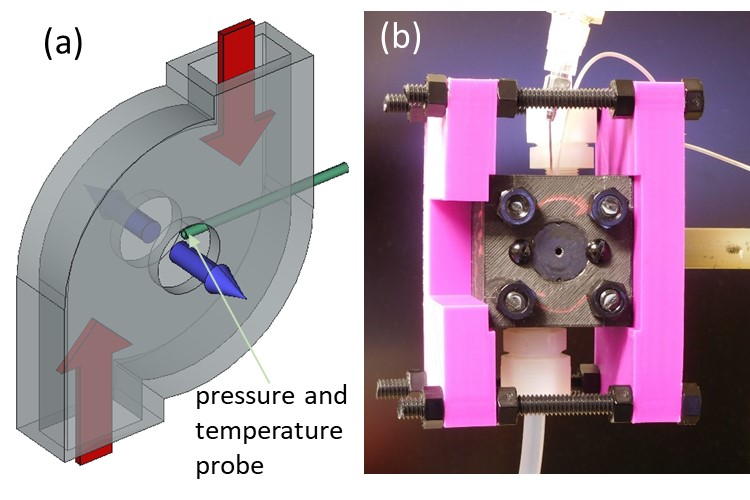}}
\caption{(a) Schematic drawing of gas lens with arrows indicating gas flow direction, and (b) an experimental device.}
\label{fig1}
\end{figure}

\begin{figure}[htbp]
\centering
\fbox{\includegraphics[width=0.5\linewidth]{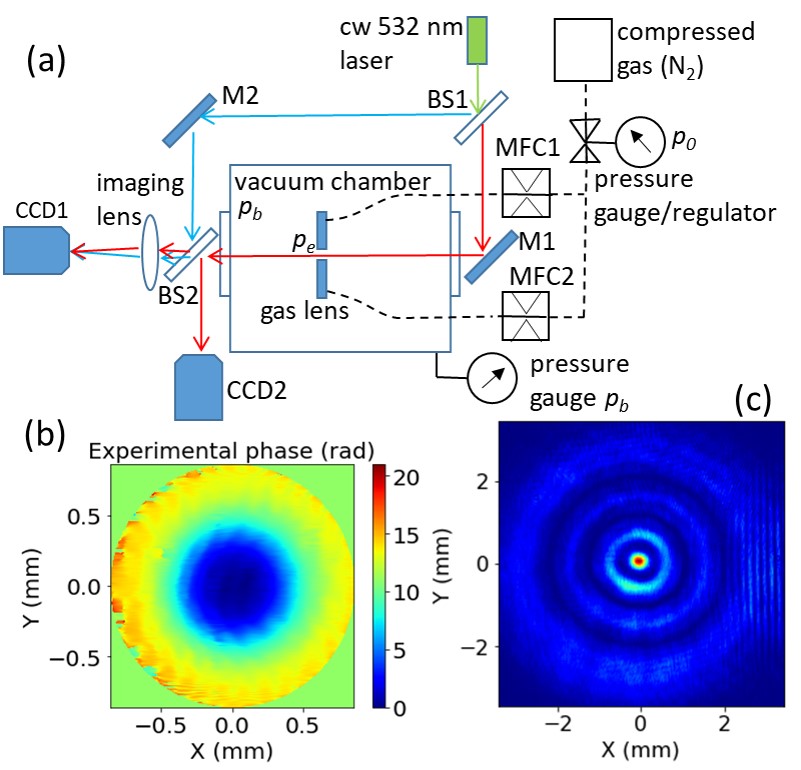}}
\caption{(a) Experimental setup. M - mirror, BS - beam splitter, MFC - mass flow controller. $p_0$ - upstream stagnation pressure, $p_e$ - gas lens exit pressure, $p_b$ - vacuum chamber backing pressure.  (b) Typical phase map derived from the interference pattern acquired by CCD1. (c) False-color, near-field intensity profile of the laser beam measured in (b) after 30 cm propagation acquired by CCD2.}
\label{fig2}
\end{figure}

The lens was secured on a holder (pink plates in Figure \ref{fig1}b) positioned in a vacuum chamber with optical axis aligned along a probe beam of a Mach-Zehnder interferometer using a 532 nm CW diode laser. A lens was used to image the exit plane of the gas lens onto the interferometer camera. The wavefront phase (typical example shown in Figure \ref{fig2}b along with lineouts in Figure \ref{fig3}a) was extracted from the interference patterns for different flow conditions. In each case the near-field intensity profile of the laser beam was recorded 30 cm from the lens (Figure \ref{fig2}c). The flow rate through the gas lens was adjusted by changing the backing pressure $p_b$ in the vacuum chamber, while keeping the upstream stagnation pressure $p_0$ fixed. For sufficiently small values of $p_b$ the flow rate was clamped due to a pressure discontinuity between the gas lens exit pressure $p_e$ and $p_b$ which is indicative of choking (blue dots in Figure \ref{fig3}d). 

The phase in the central portion of Figure \ref{fig3}a resembles a parabola and corresponds to the core of the vortex as shown in Figure \ref{fig3}b. The core has the optical properties of a negative lens with a focal length inversely proportional to the curvature of the parabola. Outside of the core the phase (and the refractive index) turns over. This localized part of the refractive index variation is responsible for the ring structure observed in the near-field image in Figure \ref{fig2}c (see Section 3c).

Studies of the longer, guiding channel vortices were made with the same experimental apparatus and with the same set of diagnostics. In this case the gas density inside the vortex was reduced, to decrease the accumulated phase shift and to make it comparable to that obtained in the shorter gas lens structure. Since the long vortex was designed for plasma-based laser-guiding applications, it was important to estimate the absolute gas density inside the vortex channel. Interferometry, however, can provide only the information about the relative density \cite{Brandi}. To get the absolute gas density we inserted a thin probe into the gas vortex as shown in Figure \ref{fig1}a (green tube). The probe consists of a small thermocouple attached to an open end of a 0.7 mm diameter tube. The tube was connected to a manometer. The probe was positioned radially at the edge of the clear aperture and longitudinally near the mid-plane of the vortex. The end of the tube was oriented perpendicularly to the gas flow streamlines, providing static pressure and temperature measurements. Pressure and temperature at known locations inside the vortex provided the gas density value from the ideal gas equation and were used for calibration of the interferometry data (Section 4).

\section{Analysis of gas lens experimental data}
\label{sec:anal}

The major tool for the experimental data analysis is an open source, in-house-developed, Python Zernike polynomials library 'zernike' \cite{Zernike}. The library has tools for expansion of an arbitrary wavefront into an orthogonal set of Zernike polynomials as well as reconstruction of the wavefront from known Zernike coefficients. Such a decomposition of the optical phase surface provides a general means to characterize the optical properties of the gas lens, including the focal length and aberrations \cite{Johnson}.

\subsection{Determination of focal length and vortex core aperture}
\label{sec:sub_3a}

\begin{figure}[htbp]
\centering
\fbox{\includegraphics[width=0.5\linewidth]{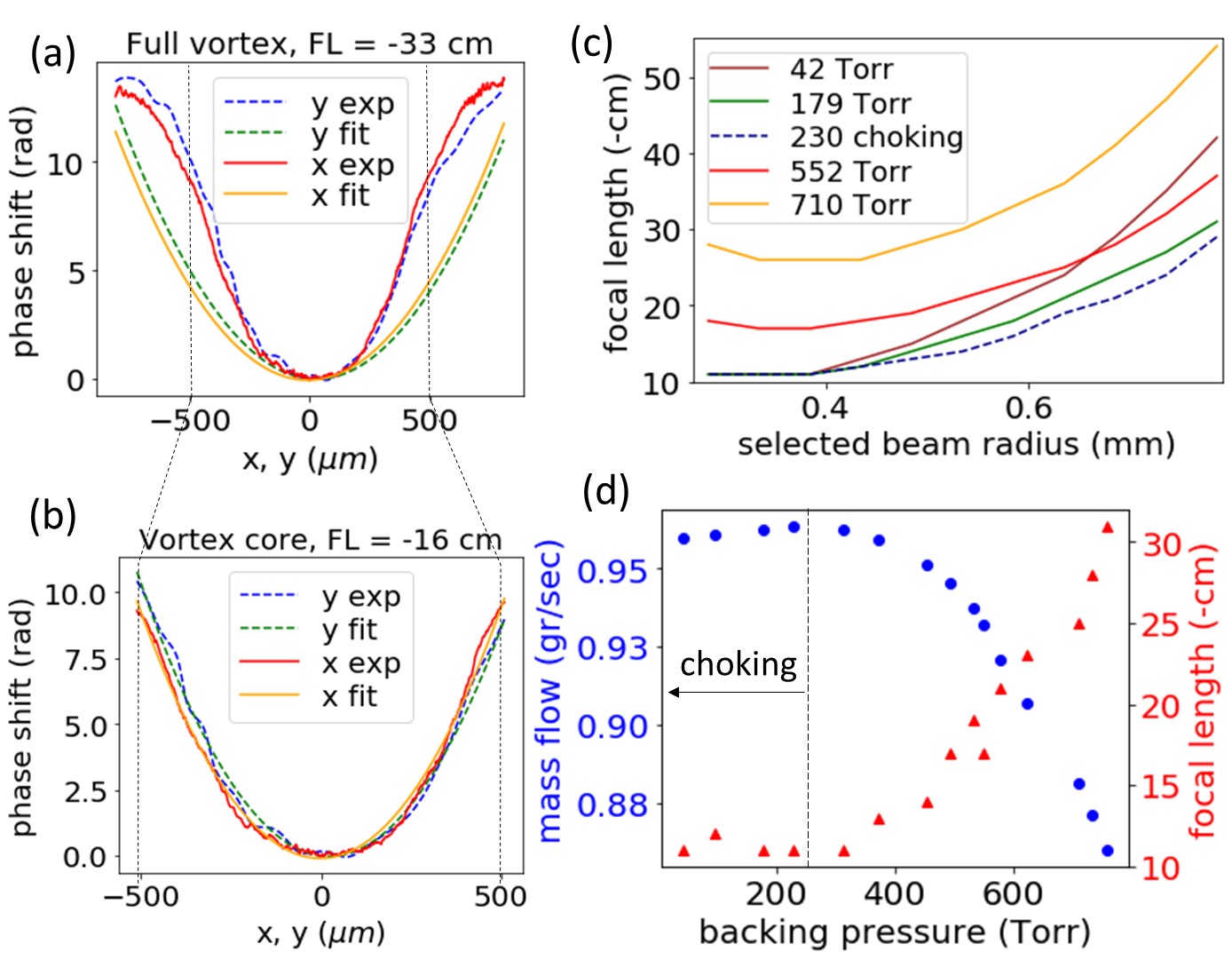}}
\caption{(a) Lineouts of the measured phase (dashed blue and solid red curves), the corresponding Zernike decomposition fit for j=1-5 (dashed green and solid yellow curves) and the focal length FL computed therefrom.  (b) The same as in (a), but truncating the data outside the black dashed lines. Inside the black dashed lines a parabola closely fits the experimental data. (c) Example of the radial scan of the phase for the core radius determination. The core radius corresponds to the largest value at which the absolute value of the focal length is nearly constant ($\approx$0.4 mm in this example). Legend in (c) shows backing pressure values. The flow was choked below 230 Torr. (d) Core (corrected) focal length (red triangles) and mass flow rate (blue dots) for different flow conditions and choking transition shown in (c).}
\label{fig3}
\end{figure}

The coefficients of Zernike polynomials derived from the experimentally measured phase provides a full two-dimensional characterization of the gas lens. The principal parameter of interest is the focal length, which is calculated using

\begin{equation}
FL=-\frac{kR^2}{4S_4},
\label{eq1}
\end{equation}
where $k=2\pi/\lambda$  is wavenumber and $\lambda$ is the wavelength, $R$ is radius of aperture of the lens, and $S_4$ is the coefficient of the $j=4$ (defocus) Zernike polynomial in standard, single-index OSA $j$ notation. The notation and the polynomials order follow that of appendix B in reference \cite{Johnson} with the modification that the single index $j$ will be used instead of the indices $n$ and $m$. As a result, the left-hand side of equation (B5) in reference \cite{Johnson} is replaced with the j-indexed coefficient $S_j$ where $j = [n(n+2)+m]/2$, and $n$, $m$ are nonnegative integers, $n \geqslant m$, and $n-m$ is even. Note that all Zernike polynomials are defined on the unit circle and are functions of the physical radius $R$. Consequently, the focal length and other optical parameters derived from the Zernike coefficients depend on the wavefront shape and in some cases on the radius $R$ (see footnote \cite{note_12}). This is clearly shown in Figures \ref{fig3}a and \ref{fig3}b, where focal length determined from Eq. \ref{eq1} changes significantly by taking different regions for decomposition of the experimental phase. The fitted curves in Figures \ref{fig3}a and \ref{fig3}b are lineouts of the reconstructed phase from the first 5 Zernike coefficients $S_1$ - $S_5$. The piston coefficient $S_0$ was selected to match the measured phase minimum at the optical axis.

The core of the gas lens vortex is expected to have a parabolic phase front (Appendix A) and its radius provides a useful aperture of the gas lens. Once the selected phase front radius is small enough and resembles an axisymmetric paraboloid, the tilt ($S_1$ and $S_2$) and the astigmatism ($S_3$ and $S_5$) Zernike coefficients become small compared to defocusing term $S_4$. The Zernike spectra of Figure \ref{fig3}a and b wavefronts are plotted side by side in Figure 6a of Appendix B. For sufficiently small radius there was little change in the value of the focal length and we used this approach to pinpoint the core radius as shown in Figure \ref{fig3}c. This figure shows scans of the focal length (Eq. \ref{eq1}) for different beam apertures and different gas flow conditions. Regardless of the flow conditions, the focal length stopped changing as the aperture radius was reduced below 0.4 mm (see footnote \cite{note_13}). This measured radius and calculated focal length defined the useful optical radius and corrected focal length of the gas lens.

\subsection{Variation of focal length versus flow conditions}
\label{sec:sub_3b}

Once we established a reliable technique for determination of the focal length and size of the vortex core, we were able to describe the focal length in terms of gas flow through the lens. The gas flow was adjusted in two different ways. First, the mass flow rate was set by the mass flow controllers MFC1 and MFC2 (Figure \ref{fig2}a), while keeping the upstream stagnation pressure $p_0$ at its highest setting of 40 PSI relative to the backing pressure $p_b$ that was set to 14.7 PSI (atmospheric pressure).  In this configuration, the corrected gas lens focal length decreased linearly with increasing mass flow rate reaching the shortest (-20 cm) value for the maximum achievable flow rate of 1.1 gr/sec through each inlet of the gas lens. 

In the second set of experiments, the mass flow controllers were fully open and operated as flow rate measuring devices, without restricting the gas flow. The upstream stagnation pressure $p_0$ was set to a constant value (30 PSI) and the mass flow rate was adjusted by changing the backing pressure $p_b$ in the vacuum chamber as described in Section \ref{sec:experiemnt}. The corrected focal length was recorded as a function of the backing pressure and is plotted in Figure \ref{fig3}d versus the measured mass flow rate. For high values of $p_b$ the focal length dependence was similar to the results from the previous paragraph, but at about 250 Torr and below the focal length and mass flow rate turned over to a constant values, indicating chocking. At the same time, the core radius of the vortex showed little dependence on the backing pressure or mass flow rate of the gas and was about 0.4 mm for the Figure \ref{fig3}c example. Once the flow choked (curves of 230, 179, and 42 Torr in Figure \ref{fig3}c), optical parameters of the core of the gas lens remained the same. 

\subsection{Beam propagation and near-field intensity distribution}
\label{sec:sub_3c}

Near-field images of the full-size probe beam propagating through the gas lens had a well-defined ring structure as shown in Figure \ref{fig2}c. Once the probe beam was down-collimated to the size of the vortex core, rings disappeared and the gas lens acted similarly to a normal defocusing lens (see Appendix B). To explain the appearance of the rings we used the open source LightPipes Python library \cite{Lightpipes}. LightPipes is a set of software tools for simulation of propagation, diffraction and interference of coherent light, and we benchmarked it by reconstructing the intensity distribution shown in Figure \ref{fig2}c from a simulated propagation of the experimental phase shown in Figure \ref{fig2}b (Appendix B).

\begin{figure}[htbp]
\centering
\fbox{\includegraphics[width=0.5\linewidth]{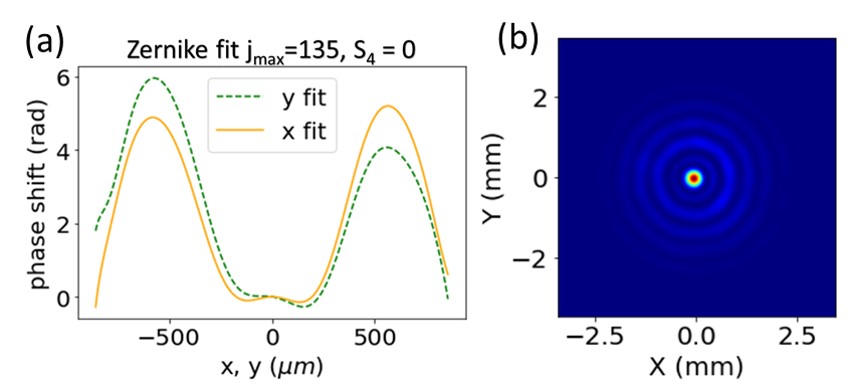}}
\caption{(a) Phase front reconstructed by summing over 135 Zernike polynomials (excluding the defocus S4 term).  The 135 Zernike coefficients are those derived from the Figure \ref{fig3}a wavefront. (b) Simulated false-color, near-field intensity distribution of this phase front after 30 cm of propagation (described in Appendix B).}
\label{fig4}
\end{figure}

To explain appearance of the rings, we calculated first 135 Zernike spectrum coefficients of the Figure \ref{fig2}b phase. The phase shift reconstructed from these coefficients was an exact replica of the experimental phase shift as shown in Figure 6b of Appendix B. Next, we excluded the defocus $S_4$ from the reconstruction. The lineouts of this wavefront are plotted in Figure \ref{fig4}a and resemble a positive ring-like lens. Not surprisingly, a similar setup, using a positive lens with a ring aperture was used in the first experimental generation of an optical Bessel beam \cite{Durin}. Simulation of propagation of such a beam using LightPipes generated Bessel-like ring structure (Figure \ref{fig4}b), except for the brighter central spot that was generated by almost undisturbed propagation of the beam center. In general, the optical structure of the spinning, neutral gas flow can be described as a combination of a negative lens and positive ring aperture lens. For practical simulation purposes a similar wavefront can be generated by combining axicon with negative lens.

\section{Gas density profile in long guiding structure }

\begin{figure*}[htbp]
\centering
\fbox{\includegraphics[width=\linewidth]{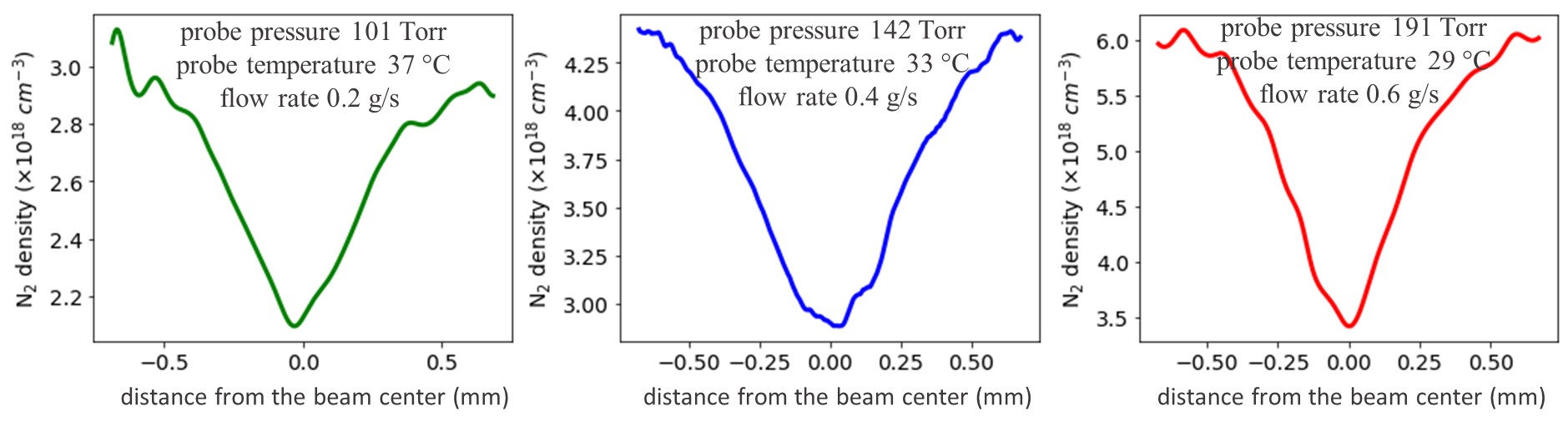}}
\caption{Calibrated gas density profiles in 16 mm long guiding structure for different flow conditions. Density was calibrated by pressure and temperature measurements with corresponding values shown in each plot.}
\label{fig5}
\end{figure*}

In contrast to short gas lenses, where the absolute value of the gas density might seem unimportant, in laser guiding applications it is critical to know the plasma (gas) density and to be able to adjust it for particular experimental requirements \cite{Kaganovich_99}. Fluid dynamics simulations of the gas lens \cite{Johnson} indicate that the radial gas density profiles are almost the same along the optical axis inside the lens. The gas density drops quickly as soon as gas exits the lens. Using these assumptions on the gas density distribution, we can retrieve the actual density values from interferometric measurements. If we ignore small contributions from the vortex structure outlets, the density is actually given by the average density measured by means of interferometry. This is especially true for the long vortices we propose to use as laser guiding structures.

The radial gas density profile in the vortex is proportional to the measured phase $\Phi (r)$

\begin{equation}
N_g(r) = \frac{N_{g0}\Phi(r)}{(n_0-1)kL},
\label{eq2}
\end{equation}
where $n_0$ is the refractive index of the gas at a specific number density $N_{g0}$, $k$ is the laser wavenumber, and $L$ is the length of the vortex \cite{Brandi}. The relative phase was extracted from the measurements of the optical interference pattern generated by the group velocity delay between a reference laser pulse and a laser pulse propagating through the vortex. The final phase change $\Phi (r)$ was calculated by subtracting this relative phase from the reference (flat) one, acquired in a separate measurement.

Since the Mach-Zehnder interferometer is highly susceptible to shot-to shot instabilities Equation \ref{eq2} required an additional calibration. The gas density calibration was done by direct determination of the gas density at one fixed location. Using the probe described in Section \ref{sec:experiemnt}, we measured gas temperature and static pressure at the edge of the clear aperture of the guiding tube (a small shadow was visible on a side of interferometry image). We assumed that small tube of the probe did not disturb the flow (we detected no effect on interference image). For known pressure and temperature the gas number density can then be calculated from the ideal gas equation:

\begin{equation*}
N_{g1} = \frac{P_1 N_A}{R_g T_1},
\end{equation*}
where $P_1$ and $T_1$ are measured by the probe pressure and temperature, $N_A$ is Avogadro number, and $R_g$ is gas constant. 

The probe was installed in a 16 mm long guiding structure and the gas density was recorded for different flow rates. Figure \ref{fig5} shows readings of the probe used in calibration of the gas density profiles. The density curves in Figure \ref{fig5} are radial lineouts of stationary vortices at corresponding flow rates. The vortices demonstrate good stability, but retrieval of the phase was hampered by fluctuations around central point and the far wings, where interferometry readings were noisy. As expected, for lower absolute gas densities the relative depth of the density depression decreases.

\section{Prospects for ionization and guiding laser beams over long distances}

While the vortex lens can find applications in both negative (neutral gas) and positive (plasma) configurations \cite{Kaganovich_18}, laser guiding is possible only in a plasma channel with a density minimum on axis. As shown above, the gas density in the vortex is already optimized for laser guiding and it is important to ionize the gas in a 'soft' manner, without significant changes to the density profile. Laser induced ionization and simple capacitive discharges modify the density by shock waves that move gas or plasma \cite{Kaganovich_14}. A similar setup involving bulk, soft ionization is under consideration at the AWAKE project at CERN \cite{Awake}. A promising approach is to use a high power helicon discharge, capable of generation of a meter-long cylinder of high density plasma \cite{Buttensch}. Ionization of the flowing gas is not well understood and in case of rotational motion might lead to new interesting physics, due to the unusual thermal properties of spinning gases \cite{Geyko_13, Geyko_16}.

The neutral gas lenses have small useful aperture (the core) and long focal length, leading to effective f-numbers in the range of hundreds. Ionization, in addition to flipping the sign of the focal length, will shorten the focal length by an order of magnitude or more \cite{Kaganovich_18} and reduce the f-number to useful values. Such a plasma lens can be used for enhanced focusing of high intensity lasers \cite{Palastro_15} and for coupling of intense laser beams into guiding channels.

In the case of very short f-number plasma lenses, the paraxial approximation breaks down, and higher order terms must be added to the radial density function to correct for spherical aberration.  Ideally one would like to match the profile of the ideal plasma lens \cite{Gordon_18}, which accounts for both radial and longitudinal density variations.  A further dimension of control is likely needed to approach this ideal, e.g., additional inlets or outlets designed to influence the longitudinal component of the flow directly. 

The parabolic gas density profile demonstrated in the guiding structures can reach the low $10^{18}$ $cm^{-3}$ gas densities (Figure \ref{fig5}). For practical laser wakefield acceleration application one requires lower densities, approaching $10^{17}$ $cm^{-3}$, over a one-meter long guiding plasma channel \cite{Leemans}. The lower gas densities require redesign of the vortex structure, while meter-scale length might be achieved by stacking multiple short stages together.

In summary, we have demonstrated robust operation of gas lens at different flow parameters. The useful aperture, corresponding to the vortex core was measured and characterized in terms of Zernike polynomials. The focal length of the gas lens can be easily adjusted by varying the flow rate.  The shortest focal length appeared at the choking condition. We generated a 16 mm long guiding structure, measured its absolute gas density distribution, and demonstrated the possibility of adjusting this density in range of $10^{18}$ $cm^{-3}$.

\section*{Appendix A. Radial gas density variation in vortex core.}
\label{sec:appendA}

Consider the flow in the straight pipe conveying the gas to the two outlets.  The fluid equations describing the gas dynamics are expressible in the form \cite{Landau}

\begin{equation}
\frac{d\rho}{dt}+\rho  \nabla \cdot \mathbf{v} = 0,
\tag{A1}
\label{eqA1}
\end{equation}

\begin{equation}
\frac{d\mathbf{v}}{dt}=-\frac{1}{\rho}\nabla p +\nu \nabla^2 \mathbf{v}+\left(\frac{\zeta}{\rho}+\frac{1}{3}\nu\right)\nabla \nabla \cdot \mathbf{v},
\tag{A2}
\label{eqA2}
\end{equation}
where $d/dt=\partial/\partial t+\mathbf{v} \cdot \nabla$, $\mathbf{v}$ is the velocity, $\rho$ is the density, $p$ is the pressure, and $\nu$ and $\zeta$ are the kinematic viscosity and the second viscosity, respectively.

Next employ the definition of vorticity $\pmb{\omega}\equiv\nabla\times\mathbf{v}$ and take the curl of both sides of Eq. \ref{eqA2} to obtain

\begin{equation}
\frac{d\pmb{\omega}}{dt}=\left(\pmb{\omega} \cdot \nabla \right) \mathbf{v} - \pmb{\omega}\left(\nabla \cdot \mathbf{v} \right) + \frac{1}{\rho^2}\nabla \rho \times \nabla p + \nu\nabla^2\pmb{\omega}+\nabla \times \left[ \left(\frac{\zeta}{\rho}+\frac{1}{3}\nu\right)\nabla \nabla \cdot \mathbf{v}\right].
\tag{A3}
\label{eqA3}
\end{equation}
Assuming $\nabla \cdot \mathbf{v}=0$ along with azimuthal symmetry $(\partial/\partial\varphi=0)$, one can expand

\begin{equation}
\mathrm{v}_r\approx-\frac{1}{2}\alpha r,
\tag{A4}
\label{eqA4}
\end{equation}

\begin{equation}
\mathrm{v}_z\approx\alpha z,
\tag{A5}
\label{eqA5}
\end{equation}
where $( r, \varphi, z )$ are the cylindrical coordinates and $\alpha$ is a constant. The streamlines obtained from Eqs. \ref{eqA4} and \ref{eqA5} are consistent with those derived from the numerical results in \cite{Johnson} near the midplane ($z \ll$ pipe length) and close to the axis $(r\to 0)$. Note also that axial flow velocity $\mathrm{v}_z\approx\alpha z$ is outward on either side of the midplane $z=0$.

Another consequence of the assumption $\nabla \cdot \mathbf{v}=0$ is that Eq. \ref{eqA3} simplifies to

\begin{equation}
\frac{d\pmb{\omega}}{dt}=\left(\pmb\omega\cdot\nabla\right)\mathbf{v}+\nu \nabla^2\pmb{\omega}+\frac{1}{\rho^2}\nabla \rho \times \nabla p.
\tag{A6}
\label{eqA6}
\end{equation}
The last term in Eq. \ref{eqA6} is referred to as the baroclinic term. Hereinafter it is assumed that the flow is isentropic, ie, $p/\rho^{\gamma}=C$ where $\gamma$ is the ratio of specific heats and $C$ is the constant associated with the Poisson adiabatic. This is the limit of barotropic flow characterized by $\nabla\rho\times\nabla p=0$. By symmetry one may assume that $\pmb{\omega}=\omega(r)\hat{e}_z$, in which case Eq. \ref{eqA6} integrates to

\begin{equation*}
\omega(r)=const.\; e^{-r^2/r^2_B},
\end{equation*}
where $r_B=2\sqrt{\nu / \alpha}$, and then

\begin{equation}
\mathrm{v}_{\varphi}=\frac{\Gamma}{2\pi r}\left(1-e^{-r^2/r^2_B} \right),
\tag{A7}
\label{eqA7}
\end{equation}
is the azimuthal velocity of the flow and $\Gamma$ is a constant (proportional to the circulation).

Equations \ref{eqA4}, \ref{eqA5} and \ref{eqA7} constitute the Burgers vortex \cite{Burgers} with the following well-known characteristics.

i) The circulation (a macroscopic measure of vorticity) along a closed contour of radius $R$ and normal to the flow is

\begin{equation*}
\oint d \mathbf{l}\cdot \mathbf{v} = \int d \mathbf{f}\cdot \pmb{\omega}=\Gamma\left(1- e^{-r^2/r^2_B}\right).
\end{equation*}

ii) The angular velocity about the $z$ axis is

\begin{equation}
\frac{\mathrm{v}_{\varphi}}{r}=\frac{\Gamma}{2\pi r^2}\left(1- e^{-r^2/r^2_B}\right)=\frac{\Gamma}{2\pi}
\begin{cases}
\frac{1}{r^2_B}, & r\ll r_B \\
\frac{1}{r^2}, & r\gg r_b
\end{cases}.
\tag{A8}
\label{eqA8}
\end{equation}
In the core region $r\ll r_B$ the motion approximates that of a rigid rotor with angular frequency $\Gamma/2\pi r^2_B$ . On the other hand, far away from the core $r\gg r_B$ the rigid rotor motion is dissipated away.

The density variation in the core region (where viscosity is unimportant) may be obtained as follows. Substituting the velocity components given in Eqs. \ref{eqA4}, \ref{eqA5} and the top line in Eq. \ref{eqA8} into Eq. \ref{eqA2} one obtains for steady flow

\begin{equation}
-\frac{1}{\rho}\frac{\partial p}{\partial r}= \frac{\alpha^2r}{4}-\left(\frac{\Gamma}{2\pi r^2_B} \right)^2 r.
\tag{A9}
\label{eqA9}
\end{equation}
In the limit of a strong vortex $\Gamma \gg 4\pi \nu$ (ie, relatively large azimuthally velocity) and making use of $p/\rho^{\gamma}=C$, Eq. \ref{eqA9} integrates to

\begin{equation}
\frac{\rho^{\gamma-1}}{\gamma-1}=\frac{\left(\Gamma/2\pi r^2_B \right)^2}{2\gamma C}r^2+const.
\tag{A10}
\label{eqA10}
\end{equation}
The $const$ is determined by employing $p_0/\rho_0=C$ at the stagnation point $(\mathbf{v}=0,\; r=z=0),$ whence

\begin{equation}
\rho (r)=\rho_0 \left[1+\frac{\gamma-1}{2}\left(\frac{\Gamma/2\pi r^2_B}{c_s}r \right)^2  \right]^{\frac{1}{\gamma-1}},
\tag{A11}
\label{eqA11}
\end{equation}
where $c_s=\sqrt{\gamma p_0/\rho_0}$ is the speed of sound. Near the axis the density has a quadratic dependence on radius

\begin{equation*}
\rho(r)\approx\rho_0 \left[1+\frac{1}{2}\left(\frac{\Gamma /2\pi r^2_B}{c_s}r\right)^2 \right],
\end{equation*}

provided $r\ll min \left[r_B, \; c_s/ \left(\Gamma/2\pi r^2_B \right)  \right]$.

\section*{Appendix B. Analysis and simulation of optical properties}
\label{sec:appendB}

\subsection*{B1. Zernike coefficients and fitting}

\begin{figure}[htbp]
\centering
\fbox{\includegraphics[width=0.7\linewidth]{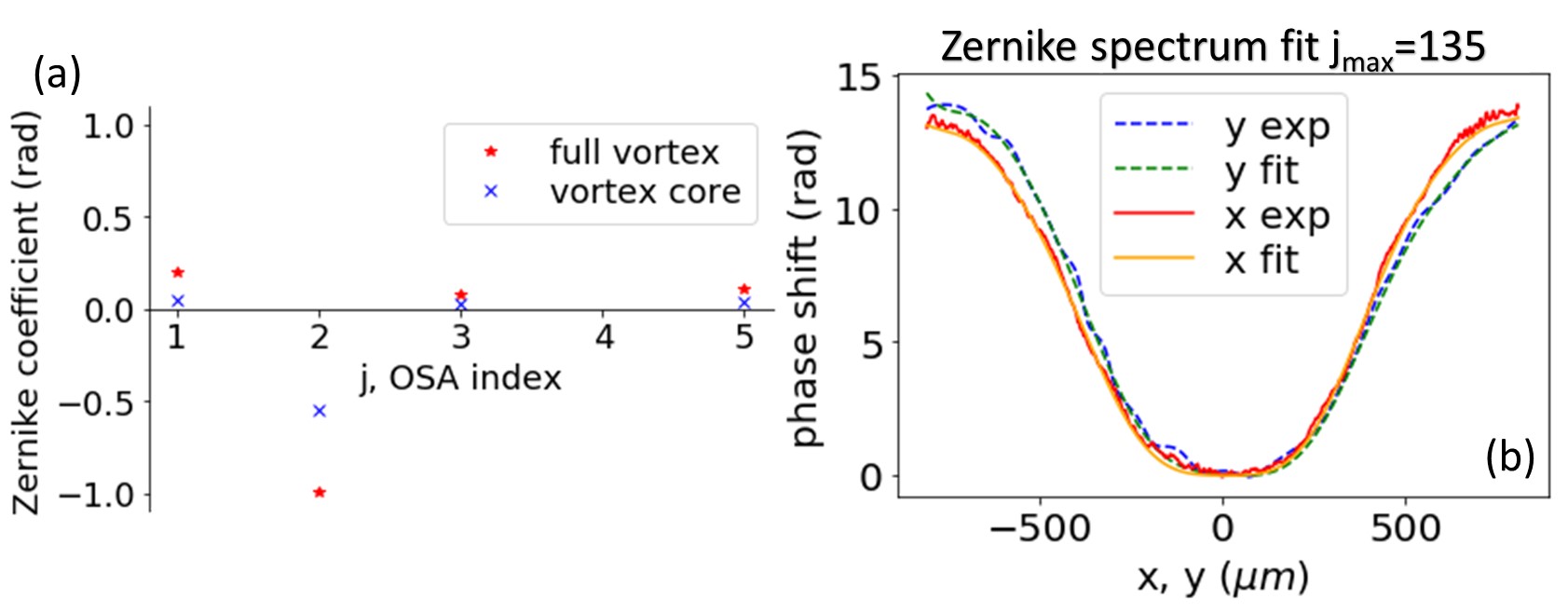}}
\caption{(a) Comparison of Zernike spectra for Figure \ref{fig3}a wavefront (full vortex) and its central part (vortex core) from Figure \ref{fig3}b. The main defocusing term $S_4$ changes from 6.6 for the full vortex to 5.9 for the vortex core and is not shown here. (b) The phase shift reconstructed from the first 135 Zernike coefficients closely matches the experimental wavefront in contrast to Figure \ref{fig3}a, where only first 5 coefficients are used.}
\label{fig6}
\end{figure}

Zernike polynomials are a complete set of orthonormal functions on the unit circle. Any phase front surface can be approximated by 'a truncated' expansion in Zernike polynomials. The first five (and all others) Zernike coefficients calculated for Figure \ref{fig3}a wavefront fittings are  the same regardless of the number of terms kept. The expansion coefficients depend on the aperture radius. Figure \ref{fig6}a shows a comparison of $S_1$, $S_2$, $S_3$, and $S_5$ Zernike coefficients derived for the wavefronts in Figures \ref{fig3}a and b. By selecting an optical aperture that matches the vortex core radius we have minimized the optical aberrations.

Zernike spectrum decompositions with more terms generate a closer match to the experimental wavefront. To get a close fit to the Figure \ref{fig3}a wavefront, we used 135 single index OSA coefficients. The corresponding function is a 15th order polynomial radially, and involves 15 Fourier modes azimuthally. The wavefront generated by this decomposition is nearly indistinguishable from the experimental phase (apart from high frequency noise at the edges) as shown in Figure \ref{fig6}b.

\subsection*{B2. Optical mode of the vortex core and lightpipes benchmarking}

\begin{figure}[htbp]
\centering
\fbox{\includegraphics[width=0.5\linewidth]{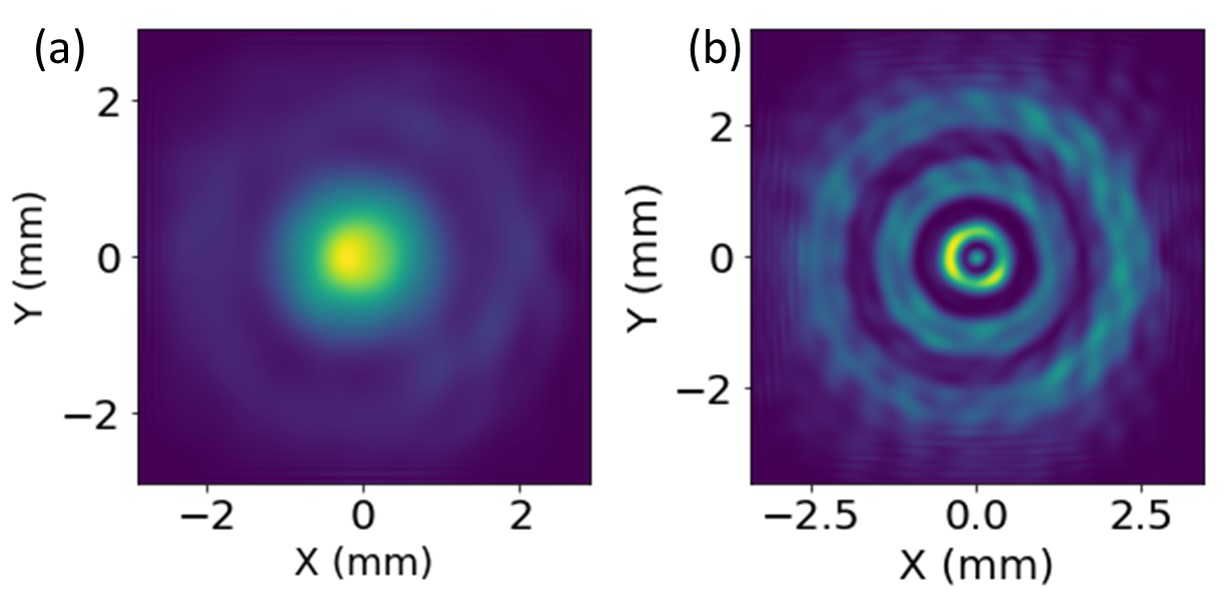}}
\caption{(a) False-color near-field image of down-collimated laser beam after propagation through the core of the vortex created by the lens used in experiments shown in Figure 3. (b) False-color near-field intensity distribution simulated with Lightpipes for laser beam after 30 cm propagation. The initial phase of the simulated beam is obtained from the Figure 2b.}
\label{fig7}
\end{figure}

The laser beam used in experiments had Gaussian radial intensity distribution with full width at half maximum (FWHM) slightly larger than clear aperture of the gas lens. The near-field intensity distribution of this beam after propagation through the lens had the ring structure described in Sections 2 and 3. To test optical quality of the core of the gas lens, the initial beam was down-collimated to about 1 mm FWHM. Using the same lens and the gas flow condition of the Figure \ref{fig3}a wavefront, we recorded the near-field image of this beam using the camera CCD2 (see setup in Figure \ref{fig2}a).  The near-field intensity distribution after 30 cm propagation is shown in Figure \ref{fig7}a.

Compared to Figure \ref{fig2}c there is no ring structure in the beam profile shown in Figure \ref{fig7}a. The weak external ring visible in Figure \ref{fig7}a was generated by the wings of the incoming Gaussian beam. These wings propagated through the gas lens outside of the vortex core.

To benchmark the Lightpipes library we simulated the near-field intensity distribution shown in Figure \ref{fig2}c from the experimental wavefront of Figure \ref{fig2}b. We hard-clipped a simulated Gaussian beam on an aperture that simulates the clear aperture of the gas lens. Then we imposed the experimental phase of the wavefront in Figure \ref{fig2}b onto the clipped Gaussian beam, inside a simulation box of 3x3 mm. The simulation box was sufficiently large to avoid interference effects on the edges during the beam propagation. Next, we simulated propagation using Lightpipes' Fresnel function over 30 cm and recorded the beam intensity distribution (Figure \ref{fig7}b).

The intensity distribution in Figure \ref{fig7}b closely resembles the one in Figure \ref{fig2}c. The dimmer central spot in Figure \ref{fig7}b is due to high sensitivity and fluctuations of the experimental phase around central point. Rings in Figure \ref{fig7}b have the same geometry and intensity distribution as in Figure \ref{fig2}c.

\section*{Funding Information}
This work was sponsored by Office of Naval Research (ONR) (N0001417WX01788) and NRL internal program.


\begin{thebibliography}{1}
\bibitem{note_12}
For example, in case of a perfect paraboloidal wavefront, the only nonzero Zernike coefficient $S_4$ will depend on selected radius $R$, whilst the focal length calculated on the basis of Eq. \ref{eq1} will be the same for any $R$.
\bibitem{note_13}
Note that for very small radii in Figure \ref{fig3}c there is slight increase in the focal length. Plots of the radial density profile show that this is due to increasing departure from a strict parabolic dependence, taking a more flat appearance near the optical axis. Such profiles are more prevalent in cases with larger apertures or lower gas flow rates.
\end{thebibliography}
\end{document}